\documentclass{aa}
\usepackage[colorlinks=true, linkcolor=blue, citecolor=blue, filecolor=blue,
                        urlcolor=blue]{hyperref}
\usepackage{graphicx,natbib,txfonts,color,amsmath,mathtools}

\bibpunct{(}{)}{;}{a}{}{,}

\begin{document}

        \title{The radial modes of stars with suppressed dipole modes}
        \author{Q.~Copp\'ee\inst{1,2} \and J.~M\"uller\inst{1,2}  
                \and M.~Bazot\inst{1} \and S.~Hekker\inst{1,2}}
        \offprints{Quentin Copp\'ee, quentin.coppee@h-its.org}  
        \institute{Heidelberg Institute for Theoretical Studies,
                Schloss-Wolfsbrunnenweg 35, D-69118 Heidelberg, Germany \and Heidelberg
                University, Centre for Astronomy, Landessternwarte, K\"onigstuhl 12,
                D-69117 Heidelberg, Germany} 
        \authorrunning{Copp\'ee et al.}
        \titlerunning{The radial modes of stars with suppressed dipole modes}
        
        
        \abstract{The \textit{Kepler} space mission provided high-quality light
        curves for more than 16 000 red giants. The global stellar oscillations extracted from these light curves carry information about the interior of the stars. Several hundred
        red giants were found to have low amplitudes in their dipole modes (i.e.~they are suppressed dipole-mode stars). A number of hypotheses (involving e.g.~a
        magnetic field, binarity, or resonant mode coupling) have been proposed to
        explain the suppression of the modes, yet none has been confirmed.} 
        {We aim to gain insight into the mechanism at play in suppressed dipole-mode stars by investigating the mode
        properties (linewidths, heights, and amplitudes) of the radial oscillation modes of
        red giants with suppressed dipole modes.}
        {We selected from the literature suppressed dipole-mode stars and compared the radial-mode properties of these stars to the radial-mode properties of stars in
        two control samples of stars with typical (i.e. non-suppressed) dipole modes.}
        {We find that the radial-mode properties of the suppressed dipole-mode
        stars are consistent with the ones in our control samples, and hence not affected by the suppression mechanism.}
        {From this we conclude that (1) the balance between the excitation and damping in radial modes is unaffected by the suppression, and by extrapolation the excitation of the non-radial modes is not affected either; and (2) the damping of the radial modes induced by the suppression mechanism is significantly less than the damping from turbulent convective motion, suggesting that the additional damping originates from the more central non-convective regions of the star, to which the radial modes are least sensitive.}

 
        \maketitle
        \section{Introduction}\label{sect:intro} The space missions CoRoT
        \citep{2006Michel} and \textit{Kepler} \citep{2010Borucki} transformed
        asteroseismology into a field of observation-driven research by providing high-quality
        light curves for more than 100 000 stellar objects, including more than 16
        000 red giants \citep{2018Yu}. This transformation is reflected in the major
        developments of red-giant asteroseismology over the last ten years
        \citep[see e.g.][for reviews]{2013Chaplin,2017H_JCD,2019GarciaBallot}. In red giants we
        observe  stochastically excited  oscillations undergoing inherently
        convective damping in the outer regions and diffusive damping
        in the core \citep[e.g.][]{1999Houdek, 2001Samadi,2009Dupret}. Advancements in the field are mostly due to the detection of non-radial
        modes in the spectrum of these stars \citep{2006Hekker,2009DeRidder}. These
        modes have a mixed character \citep{2011Beck,
        2011Bedding,2011MosserCoRoT}: they\ behave as gravity modes in the core
        region (the g-mode cavity) and as pressure modes in the outer layers of the
        star (the p-mode cavity). These mixed modes therefore carry information from both
        the p- and g-mode cavities \citep[e.g.][]{2011Beck, 2011Bedding},
        while the radial modes that are pure p-modes only carry information from
        their p-mode cavity, which makes up nearly the entire star \citep[see e.g.][]{2019Ong}.
        
        Over a decade ago, \citet{2012Mosser} identified a set of red giants with
        dipole modes with unexpectedly low amplitudes. They showed that, except for
        the dipole-mode amplitudes, the stars have similar global properties as the
        other red giants in their sample. An observational analysis of a larger sample of stars with longer light curves was performed by \citet{2016StelloPASA}. They selected red-giant-branch (RGB) stars and computed for each star the
        dipole-mode visibility (i.e.\ the ratio of the total power in the dipole
        modes to the total power in the radial modes). This measurement revealed that the suppression of the dipole modes relative to
        the radial modes decreases as stars ascend the RGB \citep[i.e.\ there is less suppression for stars with a weaker coupling between the p- and g-mode cavity; see][]{2016StelloPASA}.
        The same methodology applied to quadrupole and octupole modes showed that
        the suppression is less pronounced for higher spherical degrees \citep{2016StelloPASA}. 
 
    So far, the power in the radial modes of suppressed dipole-mode stars has not been studied. In this work we investigated whether
the energy of the radial 
    modes in suppressed dipole-mode stars is different than the energy of the radial modes in stars with dipole modes with typical visibility. We aim to gain insight into the balance between the excitation and damping of these modes as well as into the potential region where the suppression mechanism dominates (i.e. either in the outer regions or in the core) by investigating the total mode power (as per the amplitude and the height) and the mode lifetime (as per the linewidth) of the radial modes.

    One mechanism proposed to explain the suppression of the dipole modes is the presence of an internal magnetic field that dominates the regions of the star below a critical depth \citep{2015Fuller}. This depth can be linked to a critical frequency, below which the inward travelling gravity waves are refracted as magnetic waves over a broad range of spherical degrees. This means that the core magnetic field suppresses all contributions from the central regions to the observed non-radial oscillation modes (the `magnetic greenhouse effect'). The radial modes remain unaffected. The prediction for the dipole-mode visibilities of red giants made by \citet{2015Fuller} is in agreement with the observed visibilities of \citet{2012Mosser} and \citet{2016StelloPASA}. 

Using the individual frequencies of suppressed dipole-mode stars, \citet{2017Mosser} show that some of these stars still have a significant number of mixed dipole modes, albeit with lower amplitudes. This means that the central regions still contribute to the observed non-radial modes. \citet{2017Loi} propose a mechanism that induces additional damping of the non-radial modes in the presence of a core magnetic field. They show that the contribution of the central regions of the star to the non-radial modes can be partially suppressed through resonances with torsional Alfv\'en waves. These results suggest that the dipole modes can experience additional damping caused by a core magnetic field and still retain their mixed character. In the case of a strong core magnetic field, \citet{2023Rui} show that the contribution of gravity modes should be completely suppressed in red giants. They note that a more general approach (e.g. considering higher-order WKB terms or a non-harmonic time dependence) could potentially allow dipole modes to conserve a mixed character in the presence of a strong core magnetic field.
 
It is also worth noting that such strong core magnetic fields have been observed in red giants using magnetic frequency splittings \citep[see e.g.][]{2023Deheuvels}. For one of the analysed stars, the reported field is stronger than the critical field derived by \citet{2015Fuller}, suggesting that gravity waves should not be able to propagate in the central regions of that particular star. Upon analysis of its power spectrum, \citet{2023Deheuvels} found mixed-mode suppression only at low frequencies. This result is an indication that strong core magnetic fields may be linked to mixed-mode suppression.
 
    Another mechanism that can explain the suppression of dipole modes is the non-linear mode coupling between mixed modes as proposed by \citet{2019Weinberg}.  The non-radial-mode visibilities computed with this mechanism \citep{2021Weinberg} quantitatively match the visibilities of the more evolved red giants reported by \citet{2016StelloPASA}. For this mechanism as well as for the core magnetic field, the radial modes are predicted to be unaffected. 
 
    Alternatively, mode suppression could potentially be explained by the presence of a stellar companion  as its tidally induced effect can excite non-radial oscillation modes \citep[see
        e.g.][]{2013Ivanov}, indirectly impacting the properties of the
        stochastically excited oscillation modes. Statistical results also suggest that the fraction of stars with suppressed dipole modes is larger in binary stars compared to stars that are not known to be part of a binary \citep[see e.g.][]{2017Themessl}. Due to the non-radial nature of tidal effects, the radial modes are not affected by the
        tidally induced oscillations \citep[see e.g.][and references therein]{2019Beck}. 

  Finally, strong surface magnetic fields inhibit convection in the upper stellar layers, resulting in less stochastic excitation 
    and thus smaller amplitudes for radial and non-radial modes \citep[e.g.][]{2011Chaplin}.  Hence, this will have an impact on both the non-radial and radial modes, which is confirmed by \citet{2014Gaulme} and \citet{2020Schonhut-Stasik}, who found red giants in eclipsing binaries with no or reduced
    observable oscillations (i.e. the tidal effect causes enhanced magnetic activity). The presence of a surface magnetic field is the only mechanism that would impact the radial as well as the non-radial modes.

    We note here that a high core-rotation rate has also been proposed to explain the suppression of the dipole modes \citep[see e.g.][]{2014Garcia}. \citet{2014Garcia}, however, dismissed this mechanism as the cause of the suppression of the dipole modes for KIC 8561221. It is therefore most likely not the main reason for the suppression in the population of suppressed dipole-mode stars, and we do not consider it further here.
        
        In Sect.~\ref{sect:data} we specify how we selected RGB stars
        for our low dipole-mode visibility sample and our control samples, and
        in Sect.~\ref{sect:FreqAnalysis} we describe the method we applied to obtain
        the radial-mode properties of the selected stars. We define the metrics we used to investigate the radial-mode properties in Sect.~\ref{sect:RadModProp} and present the results of
        our comparisons thereof in Sect.~\ref{sect:results}. We dedicate
        Sect.~\ref{sect:discussion} to the
        discussion of our results and the conclusions.
        
        \section{Sample selection}\label{sect:data} 
        In this section we describe the selection of stars for our low-visibility sample and how we constructed the control samples. We selected two control samples based on different approaches to compare the distributions of the radial-mode properties of the suppressed dipole-mode stars to those of stars with typical, non-suppressed dipole modes.
        
        For all the stars considered in this work, we used long-cadence \textit{kepseismic}
        light curves\footnote{\url{http://dx.doi.org/10.17909/t9-mrpw-gc07}}. We only selected stars for which we have time series data with a
        time span longer than 1230 days and a filling factor larger than 0.77. These
        constraints ensure a high frequency resolution and avoid aliasing effects due to a window effect. We also cross-matched our set of stars with the APOGEE Data Release 17 (DR17)\ catalogue 
        \citep[][]{2017SDSS4,2022ApogeeDR17} to obtain stellar effective temperature $T_\text{eff}$
        and metallicity $\left[\text{Fe}/\text{H}\right]$. 
 
        We adopted the evolutionary stage (ES) provided by
        \citet{2019Kallinger} to distinguish between RGB and core-He-burning (CHeB)
        stars in our sample. If \citet{2019Kallinger} did not provide an ES determination,
        we determined the ES with the same methodology \citep{2012Kallinger,2019Kallinger}.
        This method relies on the frequencies of the three radial modes closest to the frequency of maximal
        oscillation power, $\nu_\text{max}$, and can 
        therefore be applied to all stars in our samples consistently, irrespective of their dipole-mode visibility.
        \begin{figure}[ht]
                \centering
                \includegraphics[width =\linewidth]{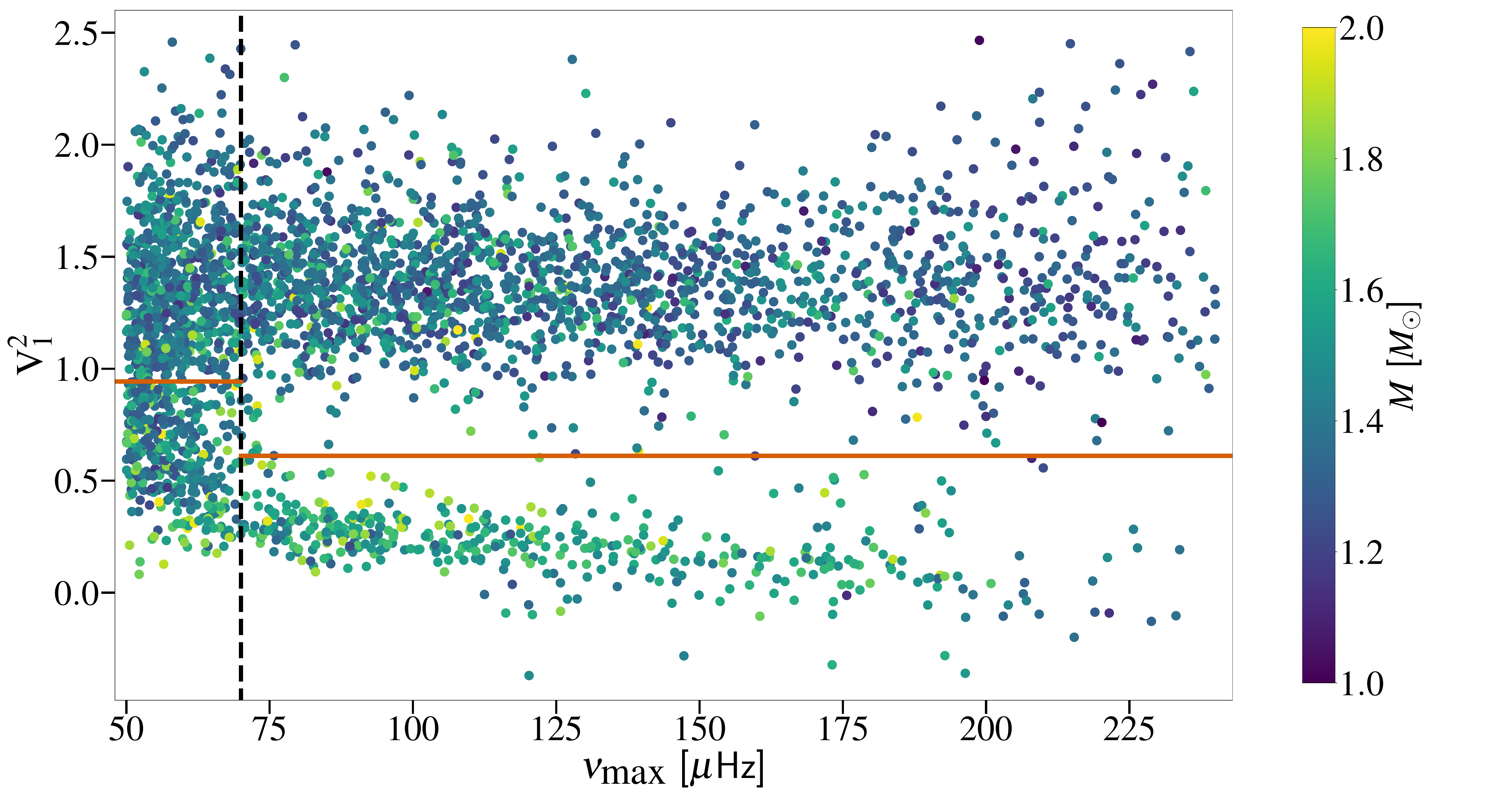}
                \caption{Dipole-visibility distribution as a function of
                $\nu_\text{max}$ colour-coded by mass \citep[all data from][]{2016StelloPASA}. The vertical dashed line indicates the boundary between the two $\nu_\text{max}$ regimes, for which the computed visibility thresholds are indicated with horizontal orange lines (see the main text
                for more details). Note that the negative visibilities are attributed to measurement scatter caused by the uncertainty in the estimation of the background model \citep[see][for further details]{2016StelloPASA}.}\label{VisDistrStello16}
        \end{figure}
        \subsection{Low-visibility sample}\label{sect:lowVisSample} 
        In the literature we find two main sources of stars with suppressed dipole modes, \citet{2012Mosser} and \citet{2016StelloNature,2016StelloPASA}. With the available data of \citet{2016StelloNature} and \citet{2017Mosser}, we selected the candidates for our sample of suppressed dipole-mode stars.
        As shown by \citet{2016StelloNature,2016StelloPASA}, the distribution in visibility of
        dipolar modes in their sample of stars is mostly bimodal: one subset of stars
        centred around 1.5 \citep[typical visibility,][]{2011Ballot} and another at
        lower visibility (see Fig.~\ref{VisDistrStello16}).
        \begin{figure*}[ht]
                \centering
                \includegraphics[width = \linewidth]{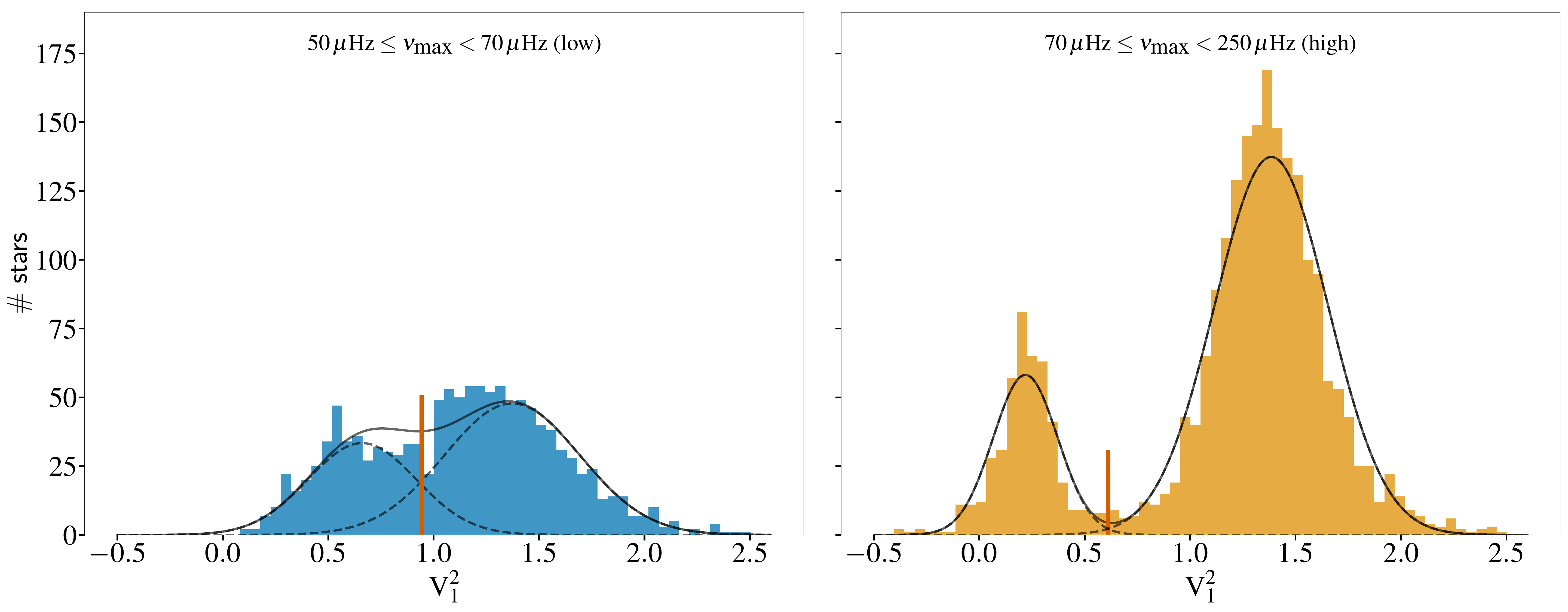}
                \caption{Dipole-mode visibility distribution in the different
                $\nu_\text{max}$ regimes for the RGB and CHeB stars analysed by \citet{2016StelloPASA}. The thresholds for the visibility are highlighted with a vertical red line. In the low $\nu_\text{max}$ regime, we selected a higher visibility threshold than that obtained from the intersection of the Gaussian components. The dipole-mode visibilities are taken from \citet{2016StelloPASA}.}\label{VisDistrNumaxRegimeStello16}
        \end{figure*}
        We divided the sample of
        \citet{2016StelloPASA} into two regimes using the frequency of maximal
        oscillation power, $\nu_\text{max}$, since the bimodality is less pronounced for the most evolved stars $\left(<70\,\mu\text{Hz}\right),$
         \begin{itemize}
             \item low, $50\,\mu\text{Hz}\le \nu_\text{max}<70\,\mu\text{Hz}$,\\
             \item high, $\nu_\text{max}\ge 70\,\mu\text{Hz}$.
         \end{itemize} 
        In each $\nu_\text{max}$ regime, we fitted a bimodal Gaussian distribution to the dipole-mode
        visibility distribution and used the intersection of the two Gaussian
        components as the threshold visibility (see Figs.~\ref{VisDistrStello16}
        and~\ref{VisDistrNumaxRegimeStello16}) and consider the stars with a dipole-mode visibility lower than these thresholds as
        candidates for our low-visibility sample. In this way, we selected in total 759 stars with low dipole-mode
        visibility from the sample of \citet{2016StelloNature,2016StelloPASA}. We further included about 40 red giants analysed by
        \citet{2017Mosser} that are not present in the sample of
        \citet{2016StelloNature,2016StelloPASA}. 
 
        After cross-matching with the stars in the APOGEE
        DR17 our sample consists of 494 red giants, 451 RGB
        and 43 CHeB stars, with low dipole-mode
        visibility and spectroscopic parameters from APOGEE DR17. Since the number of RGB stars is an order of magnitude larger than the number of CHeB stars, we focus on the RGB stars for the
        remainder of this paper.

        \subsection[]{Control samples $S_p$ and $S_c$}
        For our control samples, we pre-selected RGB stars from the APOKASC2 catalogue \citep{2019Pinsonneault} with a typical dipole-mode visibility \citep[around 1.5; see e.g.][]{2011Ballot}.
       From this, we constructed a control sample, $S_p$, of stars with observed stellar properties similar to those of suppressed dipole-mode stars. To this end, we selected
        stars by assigning to each star in our
        low-visibility sample a star with a typical dipole-mode visibility,
        similar $T_\text{eff}$, $\left[\text{Fe}/\text{H}\right]$, and $\nu_\text{max}$,
        and large frequency separation, $\Delta\nu$ (see Sect.~\ref{sect:FreqAnalysis} for details on how we obtained the last two parameters). To identify the
        most suitable configuration for our control sample $S_p$, we
        used the Kuhn-Munkres algorithm
        \citep{1955Kuhn,1956Kuhn,1957Munkres}\footnote{\textit{Python package
        munkres}, \url{https://github.com/bmc/munkres}, \copyright\,Brian M.
        Clapper}. This algorithm pairs up elements from two distinct sets and finds
        the configuration of pairs where the sum of the distances $d$ between
        elements in a pair is the smallest. We defined the distance between two
        stars in a pair as
        \begin{equation}
        \label{distance}
        \begin{split}
                d^2 = & \frac{{\left( {T_\text{eff,T}-T_\text{eff, L}}\right)}^2}{\sigma_{T_\text{eff}}^2}
                +\frac{{\left(\nu_\text{max,T}-\nu_\text{max, L}\right)}^2}{\sigma_{\nu_\text{max}}^2}\\  +&\frac{{\left(
                \Delta\nu_\text{T}-\Delta\nu_\text{L}\right)}^2 }{\sigma_{\Delta\nu}^2}+\frac{{\left(
                {\left[\text{Fe}/\text{H}\right]}_\text{T}-{\left[\text{Fe}/\text{H}\right]}_\text{L}\right)}^2}{\sigma_{\left[\text{Fe}/\text{H}\right]}^2},
        \end{split}
        \end{equation}
    where the subscript $\text{T}$ ($\text{L}$) denotes parameters of the star with typical (low) visibility.
    For the uncertainties, $\sigma_i$, we adopted the following definitions: 
         \begin{align}
             \sigma_{T_\text{eff}} &= \sqrt{\sigma_{T_\text{eff, T}, \text{intr}}^2 +\sigma_{T_\text{eff, L}, \text{intr}}^2 + 2\sigma_{T_\text{eff}, \text{emp}}^2},\\
             \sigma_{\left[\text{Fe}/\text{H}\right]} &= \sqrt{\sigma_{\left[\text{Fe}/\text{H}\right]_\text{T},\text{intr}}^2+\sigma_{\left[\text{Fe}/\text{H}\right]_\text{L},\text{intr}}^2+2\sigma_{\left[\text{Fe}/\text{H}\right], \text{emp}}^2},\\
             \sigma_{\nu_\text{max}} &= \sqrt{\sigma_{\nu_\text{max}, \text{T}}^2 + \sigma_{\nu_\text{max}, \text{L}}^2},\label{UncNum}\\
             \sigma_{\Delta\nu} &= \sqrt{\sigma_{\Delta\nu, \text{T}}^2 + \sigma_{\Delta\nu, \text{L}}^2}.\label{UncDnu}
         \end{align}
    We combined in quadrature the intrinsic uncertainties (subscript intr) of the spectroscopic parameters from APOGEE DR17 and the empirical uncertainties determined to mitigate the discrepancy between asteroseismic and spectroscopic ES classifications \citep[$\sigma_{T_\text{eff}, \text{emp}} = 44\,\text{K},\sigma_{\left[\text{Fe}/\text{H}\right],\text{emp}} = $ 0.04\,dex; see][]{2019Elsworth}. For the asteroseismic parameters, we combined in quadrature the uncertainties from our code TACO ($\sigma_{\nu_\text{max}, \text{L}}$, and $\sigma_{\Delta\nu, \text{L}}$;  see Sect. \ref{sect:FreqAnalysis} for more information) and the APOKASC2 catalogue \citep[$\sigma_{\nu_\text{max}, \text{T}}$, and $\sigma_{\Delta\nu,\text{T}}$; see][]{2019Pinsonneault}.
    
Having similar spectroscopic and asteroseismic properties does not mean that the masses of the stars
        in our samples are similar as well. To assure that a different mass 
    distribution does not affect our conclusions, we also generated a second 
    control sample, $S_c$, by randomly selecting 450 RGB stars with typical 
    dipole-mode visibilities following the cumulative density
        function (CDF) of the stellar mass distribution observed in the low
        dipole-mode visibility sample. We estimated the masses $M$ from the
        asteroseismic scaling relation \citep[see
        e.g.][]{1986Ulrich,1991Brown,1995Kjeldsen} with reference values from
        \citet{2018Themessl}.

        \section{Frequency analysis}\label{sect:FreqAnalysis} We performed the
        frequency analysis of the light curves of the stars in our samples with our
        peakbagging code TACO (Tools for the Automated Characterisation of
        Oscillations, Hekker et al., in prep.). In this section we provide an
        overview of the different steps we took to obtain the radial-mode
        properties.
 
        First we describe the power density distribution (PDS) with a global model comprising the contribution of the
        oscillations and of the background \citep[see][for more details]{2014Kallinger}. 
        We performed the remaining steps in the analysis on the background normalised PDS.

        We applied the peak detection method developed by \citet{2018GarciaSavaria} to detect the peaks in the normalised PDS. We fitted a Lorentzian function, 
        \begin{equation}
                \label{Lorentzian}
                P_\text{peak}\left(\nu\right) =  \frac{H_\text{peak}}{1+{\left(\frac{\nu - \nu_\text{peak}}{\gamma_\text{peak}}\right)}^2},
        \end{equation} to each individual detected peak 
        with $\nu_\text{peak}$, $H_\text{peak}$, and $\gamma_\text{peak}$ the central
        frequency, height, and half width at half maximum (HWHM) of the peak. We used a
        maximum likelihood estimation optimisation and computed lower limits
        of the uncertainties for each peak using a Hessian matrix. Finally, to identify the radial and quadrupole
        modes, we cross-correlated the observed normalised PDS with a synthetic normalised PDS based on the universal pattern \citep[][]{2011Mosser}.
    
        We compared the values of $\nu_\text{max}$ and $\Delta\nu$ obtained with our code to the results obtained using ABBA \citep{2019Kallinger} and the FREQ method \citep{2018Vrard}. The comparison shows that our results are in agreement with the results of ABBA and the FREQ method. For $\nu_\text{max}$ the agreement is within one $\Delta\nu$ and for $\Delta\nu$ within $ 3 \sigma$. Graphical representations of, and additional information about this comparison can be found in Appendix \ref{App:GlobalAsteroParams}.
        
        \section{Radial-mode properties}\label{sect:RadModProp} 
        The width, height and amplitude of the Lorentzian function (Eq.~\ref{Lorentzian}) describing an
        oscillation mode carry information about the excitation and damping
        processes affecting this mode \citep[see e.g.][and references therein]{2017H_JCD}. 
        
        The HWHM $\gamma_\text{peak}$ or linewidth
        $\Gamma_\text{peak}$ (full width at half maximum) of a mode are directly
        related to the mode-damping rate, $\eta,$ and the mode lifetime, $\tau,$
        \begin{equation}
                \label{ModeDamping}
                \gamma_\text{peak}= \frac{\Gamma_\text{peak}}{2} = \frac{\eta}{2\pi} = \frac{1}{2\pi\tau},
        \end{equation}while  $A_\text{peak}^2$ (the square of the mode amplitude, i.e.\,the area under the Lorentzian) and $H_\text{peak}$ (the height of the Lorentzian; see Eq.~\ref{Lorentzian}) are related to the total energy of
        the mode and are defined as
        \begin{equation}
                \label{ModeAmplitude}
                A_\text{peak}^2= \pi\gamma_\text{peak}H_\text{peak}.
        \end{equation}  
                        \begin{figure*}[ht]
                \centering
                \includegraphics[width = \linewidth]{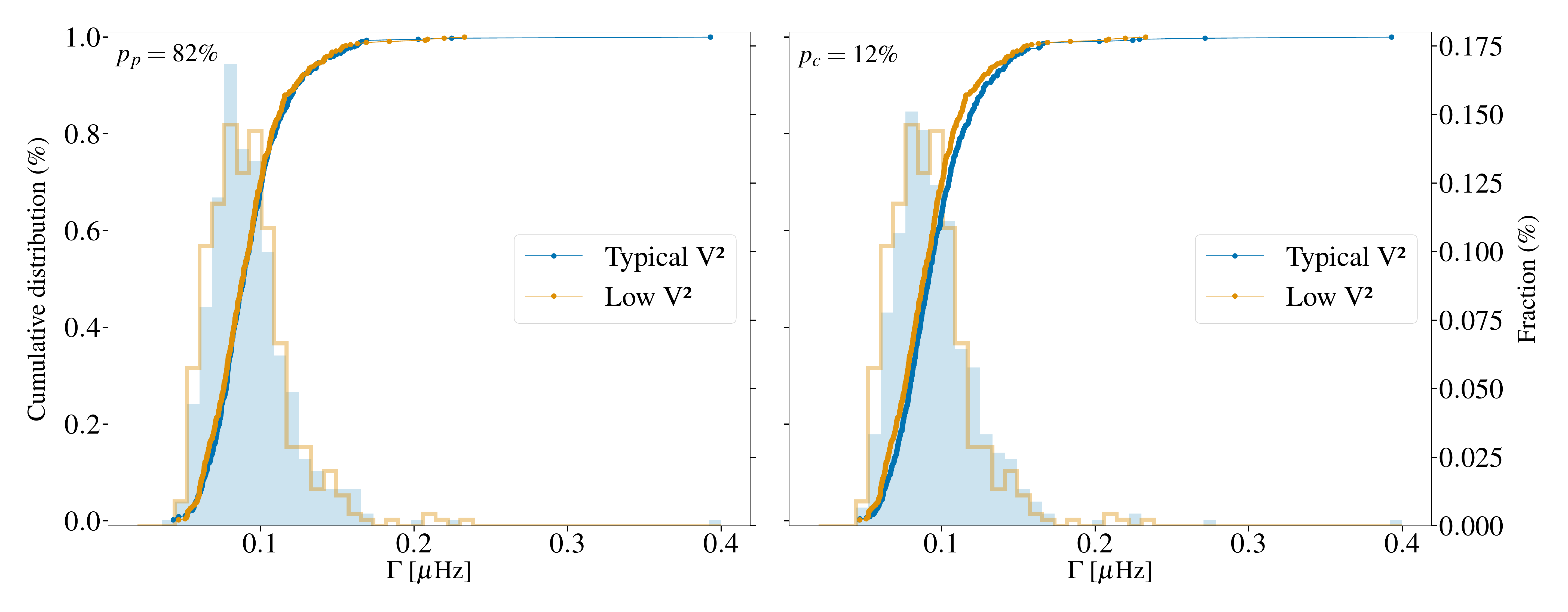}
                \caption{Cumulative distributions of the global linewidth, $\Gamma,$
                for the $S_p$ (left) and $S_c$ (right) control samples (blue)
                compared to the low-visibility sample (orange). The distributions are
                shown as histograms in the background (see the right vertical axis for the fraction of stars per bin). The respective p-values of the KS test are shown in the top-left corner of the panels.}\label{GammaTeffRGBPAIR}
        \end{figure*}
        To condense the information of the different radial modes per star into a single value for the linewidth, amplitude, and height, we defined a global
        radial linewidth ($\Gamma$), amplitude ($A$), and height ($H$) as the classical weighted average properties of the three radial
        modes closest to $\nu_\text{max}$, namely the modes for which we typically obtain the most precise mode parameters: 

    \begin{align}\label{GloBalGammaDef}
            \Gamma &\equiv \frac{w_1\Gamma_1 +w_2\Gamma_2+w_3\Gamma_3}{w_1+w_2+w_3}\\
                         A &\equiv\frac{W_1A_1 +W_2A_2+W_3A_3}{W_1+W_2+W_3} \label{GloBalADef}\\
            H &\equiv\frac{\mathcal{W}_{1}H_{1} +\mathcal{W}_{2}H_{2}+\mathcal{W}_{3}H_{3}}{\mathcal{W}_{1}+\mathcal{W}_{2}+\mathcal{W}_{3}}, \label{GloBalHDef}
    \end{align}

    \noindent with the weights $w_i= \sigma^{-2}_{\Gamma_i}$, $W_i= \sigma^{-2}_{A}$, and $\mathcal{W}_i= \sigma^{-2}_{{H}_i}$ defined as a function of the uncertainties
        $\sigma_{\Gamma_i}$, $\sigma_{A_i}$, and $\sigma_{{H}_i}$ of the $i$th radial mode.\@ To take the spectral response of the \textit{Kepler} instrumentation into
        account, we applied the bolometric correction from \citet{2011Ballot} to the
        global radial amplitude. In other words, we computed the bolometric
        amplitude $A_{\text{Bol}}$ for each star:
        \begin{equation}\label{GloBalAbolDef}
                A_{\text{Bol}} = A \cdot {\left(\frac{T_{\text{eff}}}{5934\ \text{K}}\right)}^{0.8}.
        \end{equation} 

    We compared our resulting amplitudes and linewidths with values available in the literature. We find that the individual and global radial-mode properties for the stars in our
        samples are similar to the ones computed in the same way based on the linewidths and amplitudes reported by \citet[ABBA]{2019Kallinger}. We find agreement with the results based on the data from \citet{2019Kallinger} within 10$\%$, which
        is within the typical uncertainties. Furthermore, our
        bolometric amplitudes are consistent with the results obtained by
        \citet{2018Vrard}. However, our global radial-mode linewidths are about 40\% smaller than the ones reported by \citet[i.e. the FREQ method]{2018Vrard}. By comparing the linewidths obtained by \citet{2018Vrard} and \citet{2019Kallinger}, we also observe that the linewidths in \citet{2018Vrard} are systematically broader. Although we find that our linewidths are overall narrower for the stars we have in common, the distributions in the global radial-mode parameters are still consistent independent of the datasets that were chosen (the dataset resulting from our code or the datasets from \citealt{2018Vrard} or \citealt{2019Kallinger}). Comparisons of the differences between the datasets can be found in Appendix \ref{App:GlobalRadParams}.
        
        \section{Results}\label{sect:results} 
\begin{figure*}[ht]
                \centering
                \includegraphics[width = \linewidth]{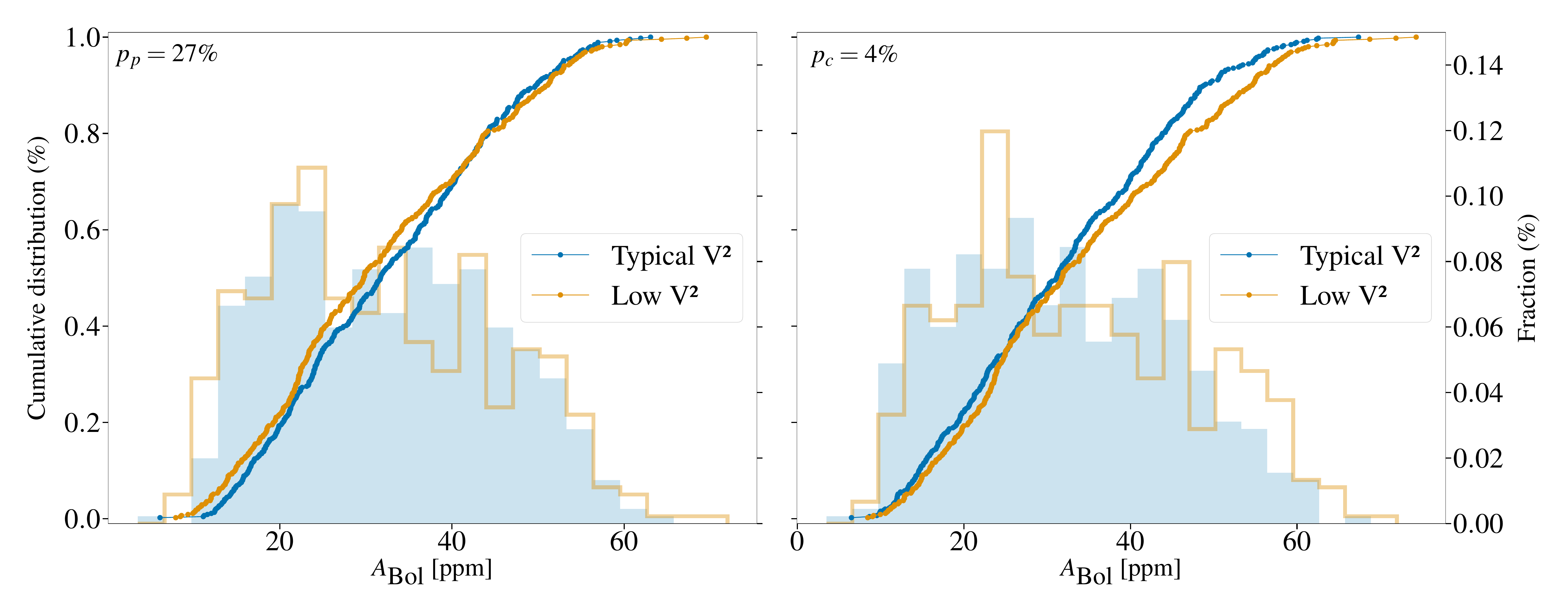}
                \caption{Same as Fig.~\ref{GammaTeffRGBPAIR}, but now for bolometric amplitude,
                        $A_\text{Bol}$.}\label{ABolLumRGBPAIR}
                \end{figure*}

\begin{figure*}[ht]
                \centering
                \includegraphics[width = \linewidth]{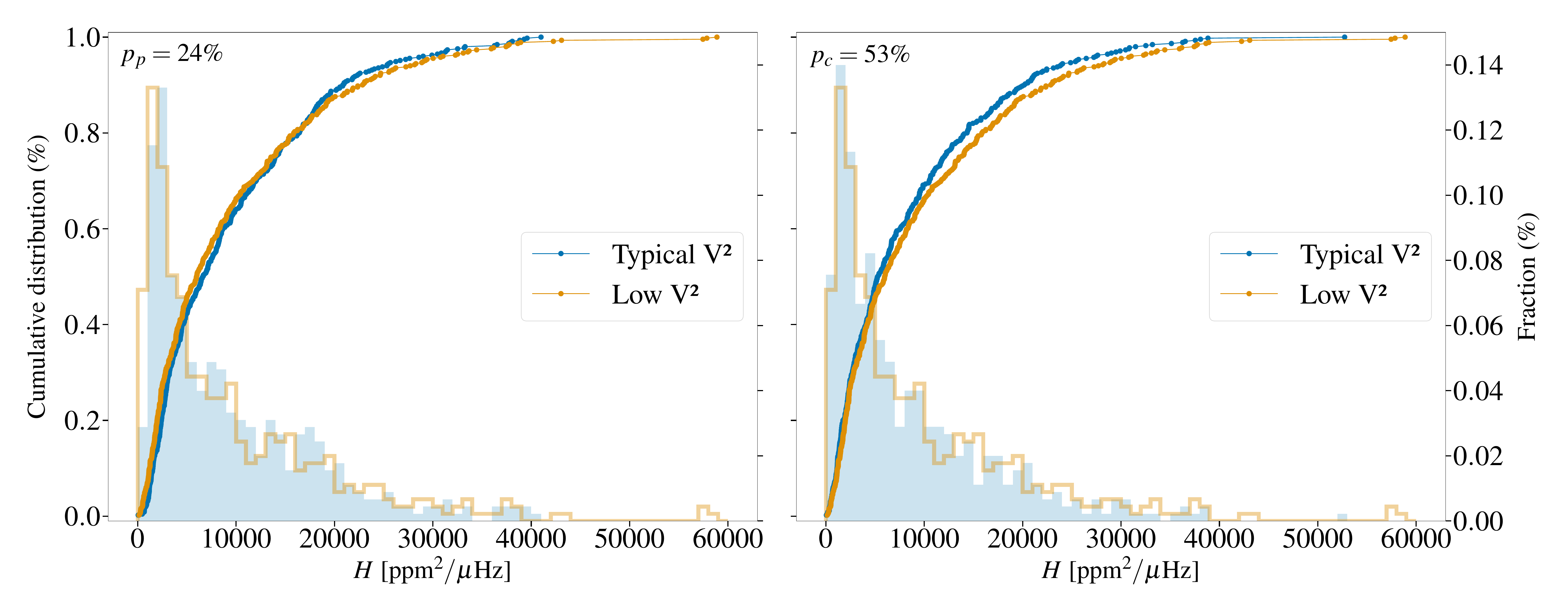}
                \caption{Same as Fig.~\ref{GammaTeffRGBPAIR}, but now for height, $H$.}\label{HRGBPAIR}
                \end{figure*}

In this section we present the comparison of the
        CDFs of global radial-mode properties
        ($\Gamma$, $H,$ and $A_{\text{Bol}}$) from each control sample to the ones from
        the low dipole-mode visibility sample (see Figs.~\ref{GammaTeffRGBPAIR},~\ref{ABolLumRGBPAIR}, and~\ref{HRGBPAIR}). For a quantitative comparison, we used the p-values of the Kolmogorov-Smirnov two-sample
        test \citep[the KS test hereafter;][]{1958Hodges}, that is,\ the probability that the two samples
        are taken from the same underlying distribution (the null hypothesis). A p-value smaller
        than 1\,\% indicates that we can reject the null hypothesis.

        We find that we cannot reject the hypothesis that the distributions in $\Gamma$ for stars in $S_p$ and $S_c$ are similar to the distribution for suppressed dipole-mode stars (see Fig.~\ref{GammaTeffRGBPAIR}, where the p-values of the KS test are 82 and 12\,\%, respectively). Since $\Gamma$ can be related to the average mode-damping rate $\eta$, and consequently to the average mode lifetime, $\tau$ (see Eq.~\ref{ModeDamping}), we conclude that the convective damping of the radial modes is
        not altered by the suppression mechanism in suppressed dipole-modes stars.
        
        We additionally find that we cannot reject the hypothesis that the distribution in the bolometric amplitude $A_\text{Bol}$ and $H$ observed in $S_p$ and $S_c$
        are similar to the distribution observed for suppressed dipole-mode stars (see
        Fig.~\ref{ABolLumRGBPAIR}, where the
        p-values of KS test are 27 and 4\,\%, respectively, and Fig.~\ref{HRGBPAIR}, where the
        p-values are 24 and 53\,\%). This indicates that the total power in the
        radial modes (scaling as $A_\text{Bol}^2$ and as $H$) is comparable for stars with low
        and typical dipole-mode visibility. 
        
        We checked that the conclusions we draw from the KS test are not influenced
        by the randomness implemented in our selection procedures. We therefore
    generated 5000 realisations of $S_p$ and $S_c$ by 
    randomly selecting a value in the uniformly distributed interval 
    $\left[x-\sigma_x, x+\sigma_x\right]$ for parameter $x$ ($T_{\text{eff}}$, $\left[\text{Fe}/\text{H}\right]$,
    $\nu_\text{max}$, $\Delta\nu$, $M$, $\Gamma$, $H$, and $A_{\text{Bol}}$) with uncertainty $\sigma_x$.
        If we cannot reject the null hypothesis of the KS test for a large majority of these realisations compared to the
        distribution observed in the
        low-visibility sample for a given parameter, we conclude that this parameter has a similar
        distribution for both control and low-visibility sample. For all our parameters we can conclude that the 
    low-visibility and control samples are drawn from the same underlying distribution.

        Additionally, we repeated the selection
        procedure to generate four additional control samples $S_c$ to investigate the 
    influence of the random selection of stars for our control sample. It is 
    important to mention that the same star can be selected for
        different samples. Again, we find no significant difference in the
        distributions of the radial-mode properties of low- and typical-visibility
        stars for any of these samples.
        
        Finally, we checked if our results are independent of the chosen metric in Eqs. \ref{GloBalGammaDef}, \ref{GloBalADef}, and \ref{GloBalHDef}. We repeated our comparison with the properties of the three central radial modes, and with the arithmetic mean instead of a weighted mean. For all metrics, we obtain the same results: the distributions of the radial-mode properties for our low-visibility stars are consistent with the properties of the stars in our control samples.
        
        \section{Discussion and conclusions}\label{sect:discussion}
        In this study we measured and compared the radial-mode properties of stars with suppressed and non-suppressed dipole modes. We find that the linewidths, and by extension the radial-mode damping rates of the radial modes (see Eq.~\ref{ModeDamping}), are similarly distributed for both types of stars.
     
        We also show that the bolometric amplitudes and the heights of the radial modes are distributed similarly for suppressed and non-suppressed dipole-mode stars. Moreover, since the radial-mode height is independent of the mode inertia, our results show that the mode energy (related to the height and the squared bolometric amplitude) is similarly distributed for both types of stars.

        As the mode energy represents the balance between damping and excitation processes \citep[see e.g.][]{2017H_JCD}, we infer from our results that the excitation of the radial modes is unaffected by the mechanism causing the suppressed-dipole modes. Assuming that modes of different spherical degrees are excited in a similar manner, this means that all modes are excited similarly in stars with suppressed and non-suppressed dipole modes. This then leads us to conclude that the observed suppression is caused by additional damping and not by lack of excitation.
        
        The additional source of damping does not significantly affect the radial modes (i.e. the convective damping in radial modes seems unaffected). Under the assumption that the convective damping of the dipole modes is similar to the convective damping of the radial modes \citep[see e.g.][]{2017Mosser}, the additional damping is likely taking place in the stellar core, to which mixed modes are more sensitive than the radial modes.
    
        Since we find that the excitation of the radial modes remains unaffected by the suppression mechanism, our results suggest that a mechanism impacting the mode excitation, such as the presence of a strong surface magnetic field, is likely not the cause of the suppression of the dipole modes in low-visibility stars. For the core magnetic field mechanism, \citet{2015Fuller} note that the magnitude of the core magnetic field is assumed to be too low to affect the radial modes. Different theoretical approaches, such as those developed by \citet{2017Lecoanet} or \citet{2017Loi}, confirm that the presence of a core magnetic field can reduce the dipole-mode amplitudes without affecting the radial modes. Additionally, \citet{2016Cantiello} extended the theoretical analysis of
        \citet{2015Fuller} by investigating the generation, evolution, and
        detectability of core magnetic fields and corroborated the results from
        \citet{2015Fuller}. Based on a more general analysis, \citet{2023Rui}
        conclude that above a given magnetic field strength, inward travelling
        gravity waves will indeed be refracted at a critical depth and will be
        converted to outgoing slow magnetic waves that dissipate before reaching the stellar surface, leaving the radial-mode properties unaffected. 
        
        The resonant mode coupling is a three-wave interaction between mixed modes where a non-radial mode destabilises two other modes with an initially low amplitude \citep{2019Weinberg}. According to \citet{2021Weinberg}, the relatively small displacements near the stellar centre and the large damping rate dominated by convective damping prevent the radial modes from being impacted by this interaction.
        For red giants in binaries, \citet{2019Beck} conclude
        that the tidal effects do not significantly impact the radial modes, as no change in the large frequency separation and no radial-mode amplitude modulation were observed. Based 
        on the binary fraction (about 10 \%) in our low dipole-mode
        visibility sample, it appears nevertheless unlikely that binarity alone can cause the suppression.
       
        In summary, our findings support either a strong core magnetic field or resonant mode coupling as the suppression mechanism, while surface-magnetic fields seem rather unlikely. According to the binary fraction in the suppressed dipole-mode sample, tidal effects are likely not the cause of the mode suppression. Further investigations into core magnetic fields will be presented in a forthcoming work by M\"uller et al. (in prep.).

        \begin{acknowledgements}
        We thank the referee for their useful comments and remarks that considerably improved the manuscript. We acknowledge funding from the ERC Consolidator Grant DipolarSound (grant agreement \# 101000296). 
        
        \end{acknowledgements}
        \bibliographystyle{aa.bst}
        \bibliography{allreferences}
 
        \begin{appendix}
 
        \section{Comparison of $\nu_\text{max}$ and $\Delta\nu$ with literature values}\label{App:GlobalAsteroParams}
\begin{figure*}[ht]
                \centering
                \includegraphics[width =0.9 \linewidth]{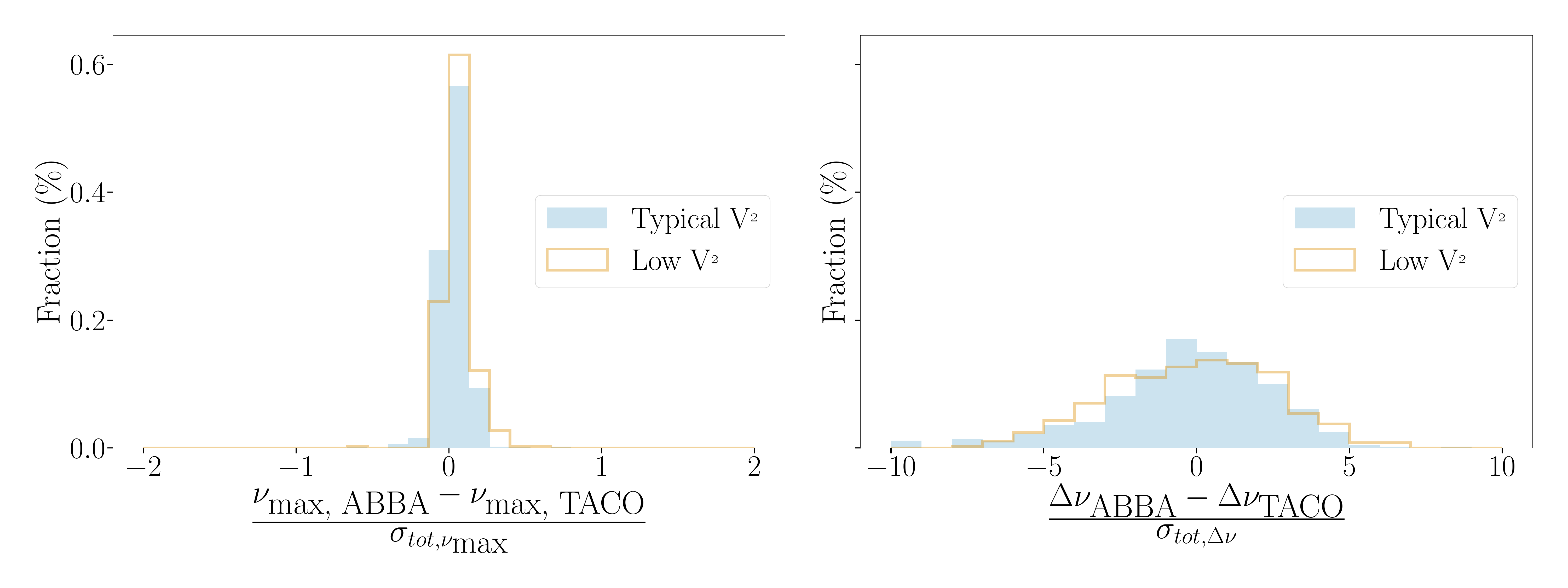}
                \caption{Difference between the $\nu_\text{max}$ (left) and $\Delta\nu$ (right) derived in this work and the results of ABBA \citep{2019Kallinger} expressed in $\sigma$ for stars in the low-visibility sample (orange) and $S_c$ (blue).}\label{RelDiff_Num_Dnu_TACO_ABBA_CUM}
        \end{figure*}

\begin{figure*}[ht]
                \centering
                \includegraphics[width =0.9 \linewidth]{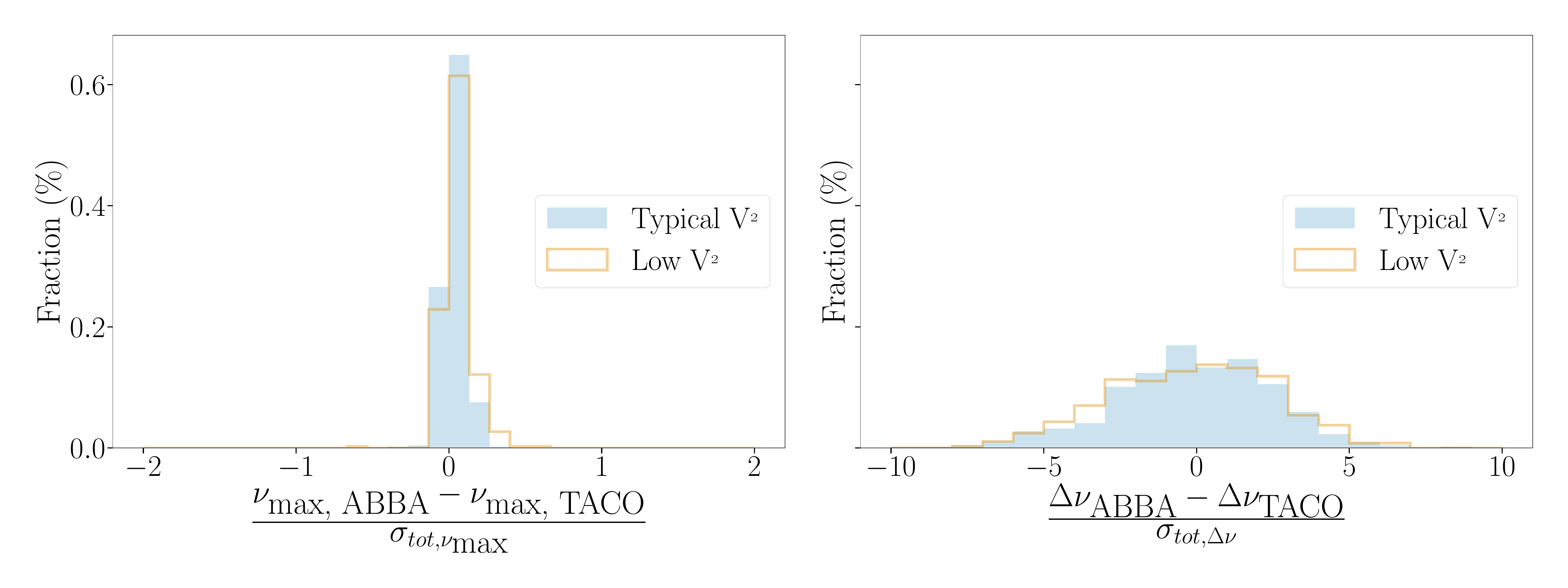}
                \caption{Same as in Fig. \ref{RelDiff_Num_Dnu_TACO_ABBA_CUM}, but now for $S_p$ (blue).}\label{RelDiff_Num_Dnu_TACO_ABBA_PAIR}
        \end{figure*}

\begin{figure*}[ht]
                \centering
                \includegraphics[width =0.9 \linewidth]{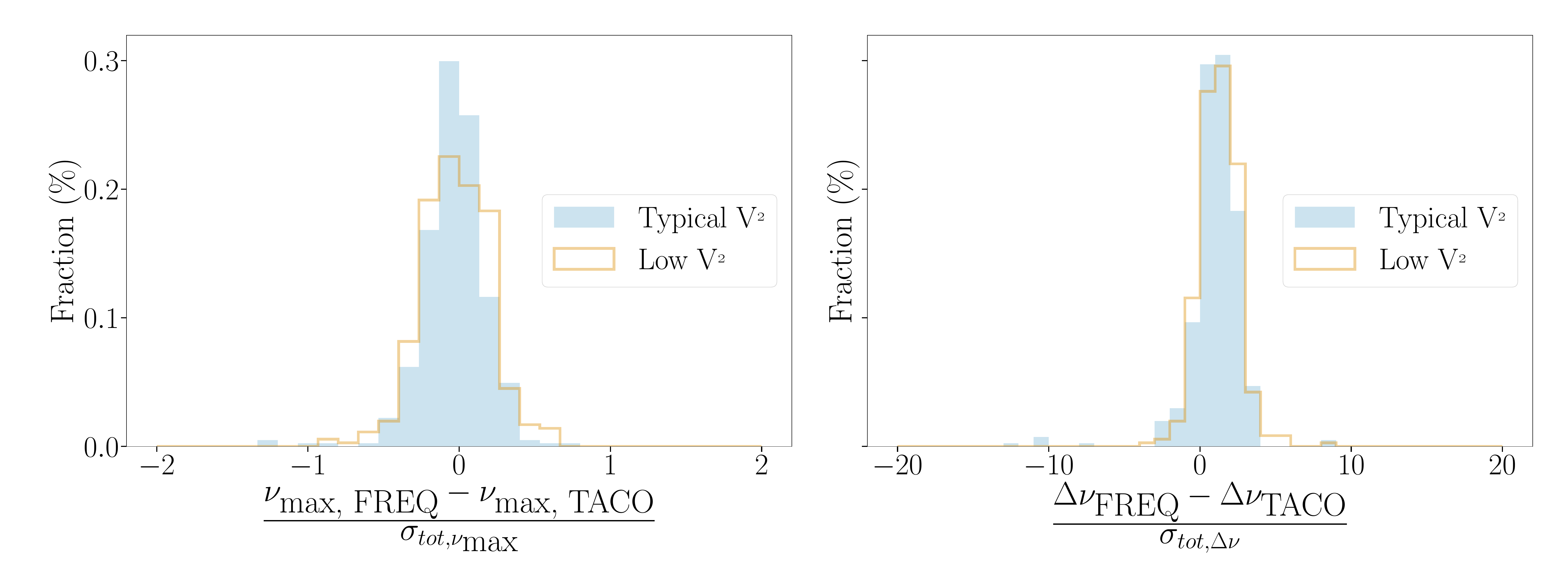}
                \caption{Same as in Fig. \ref{RelDiff_Num_Dnu_TACO_ABBA_CUM}, but now for a comparison with the FREQ results \citep{2018Vrard}.} \label{RelDiff_Num_Dnu_TACO_FREQ_CUM}
        \end{figure*}

\begin{figure*}[ht]
                \centering
                \includegraphics[width =0.9 \linewidth]{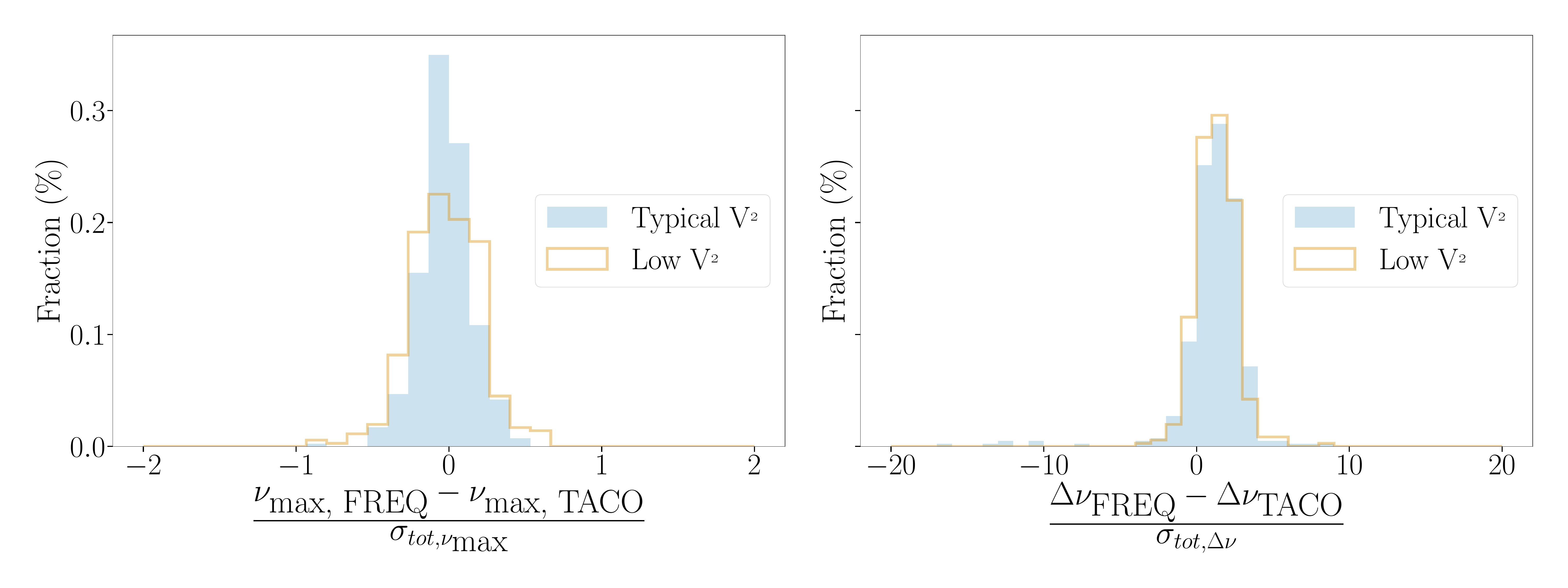}
                \caption{Same as in Fig. \ref{RelDiff_Num_Dnu_TACO_FREQ_CUM}, but now for $S_p$ (blue).}\label{RelDiff_Num_Dnu_TACO_FREQ_PAIR}
        \end{figure*}

        In this section we compare the values of $\nu_\text{max}$ and $\Delta\nu$ obtained with our code to the values found in the literature. We find agreement between our results and the data from \citet{2018Vrard} and \citet{2019Kallinger} for the stars we have in common with their samples (about 70\% of our samples).
         We show in Fig. \ref{RelDiff_Num_Dnu_TACO_ABBA_CUM} and \ref{RelDiff_Num_Dnu_TACO_ABBA_PAIR} as well as in Fig. \ref{RelDiff_Num_Dnu_TACO_FREQ_CUM} and \ref{RelDiff_Num_Dnu_TACO_FREQ_PAIR} the difference in $\nu_\text{max}$ and $\Delta\nu$ between our results for our low-visibility sample, $S_c$ and $S_p$ and the data published by respectively \citet{2019Kallinger} and \citet{2018Vrard} as a function of their uncertainties. We defined the uncertainty on the difference in $\nu_\text{max}$ as one $\Delta\nu$ and on the difference in $ \Delta\nu$ as the combination of the individual uncertainties in quadrature (similar to e.g. in Eq. \ref{UncDnu}). We find that our values for $\nu_\text{max}$ are within 1 $\Delta_\nu$ and  $\Delta_\nu$ within $ 3 \sigma$ of the given uncertainties. The few outliers in the differences of $\Delta\nu$ are caused by a different number of detected radial modes.

        \section{Comparison of individual linewidths, amplitudes, and heights with literature values}\label{App:GlobalRadParams}
\begin{figure*}[ht]
                \centering
                \includegraphics[width=0.9\linewidth]{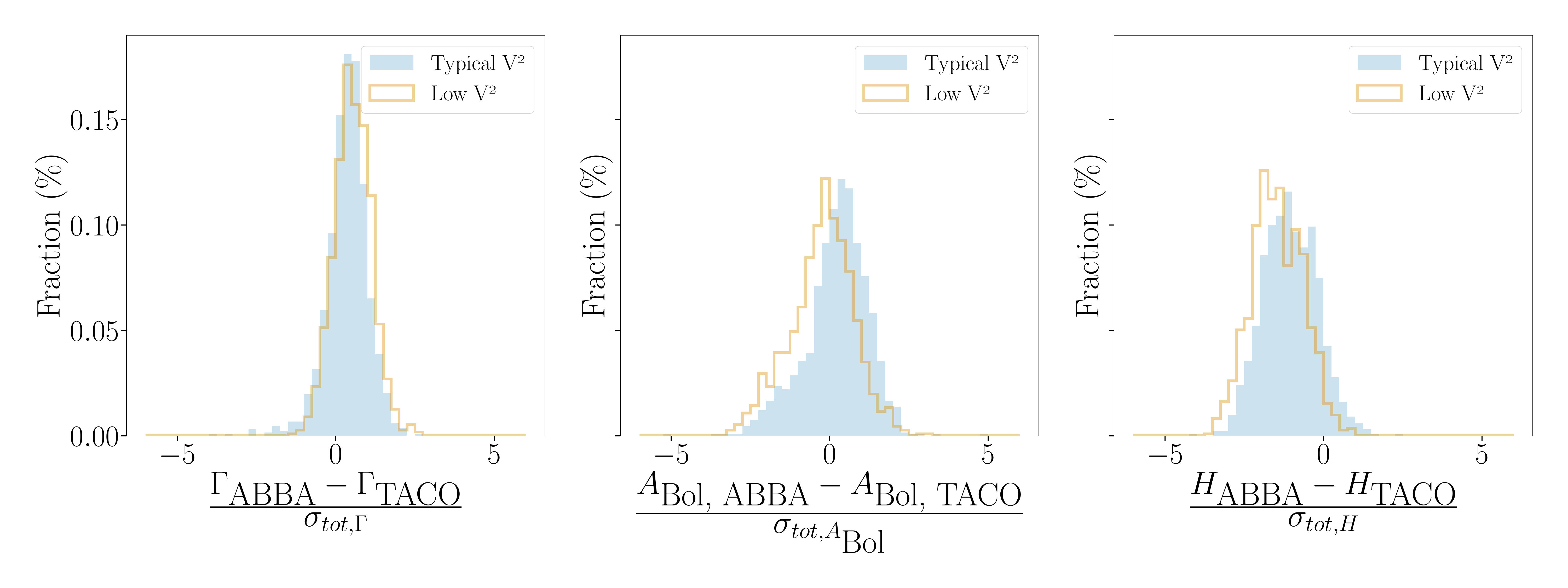}
                \caption{Difference between the linewidths (left), bolometric amplitudes (middle), and heights (right) based on our results and the \citet{2019Kallinger} data for stars in the low-visibility sample (orange) and $S_c$ (blue).}\label{RelDiff_Gamma_ABol_TACO_ABBA_CUM}
        \end{figure*}

\begin{figure*}[ht]
                \centering
                \includegraphics[width =0.9\linewidth]{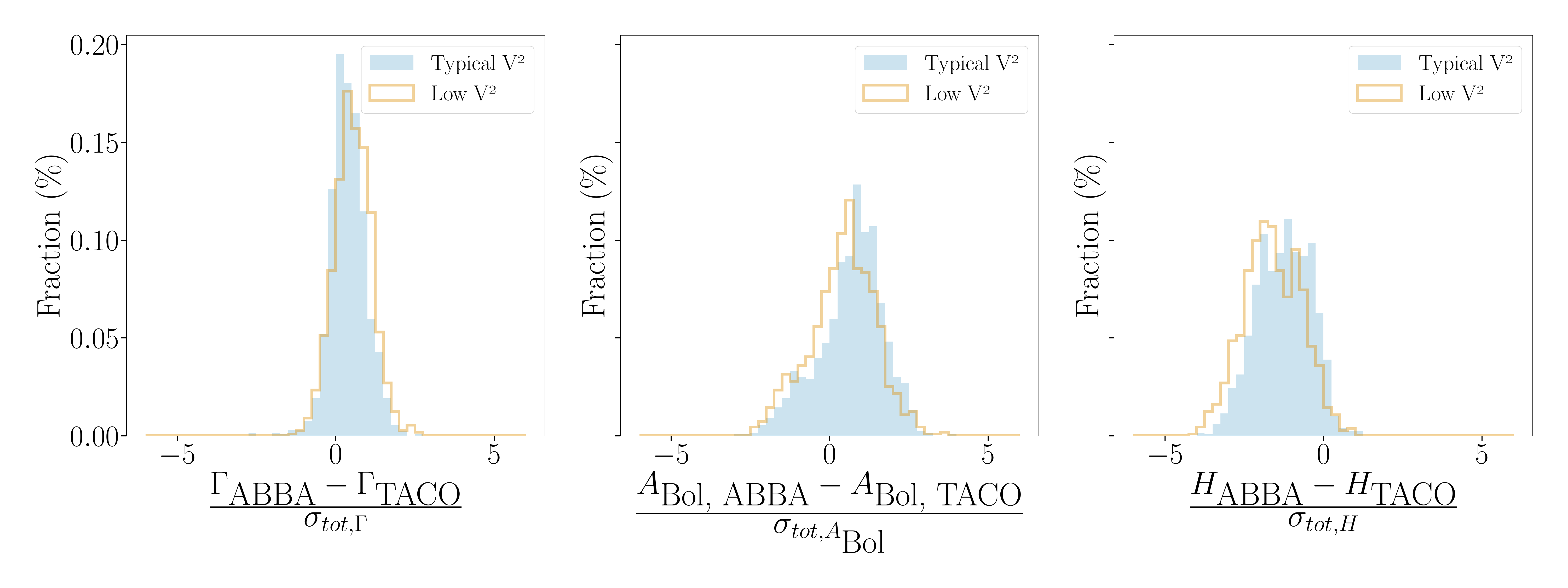}
                \caption{Same as in Fig. \ref{RelDiff_Gamma_ABol_TACO_ABBA_CUM}, but now for $S_p$ (blue).} \label{RelDiff_Gamma_ABol_TACO_ABBA_PAIR}
        \end{figure*}

\begin{figure*}[ht]
                \centering
                \includegraphics[width =0.9 \linewidth]{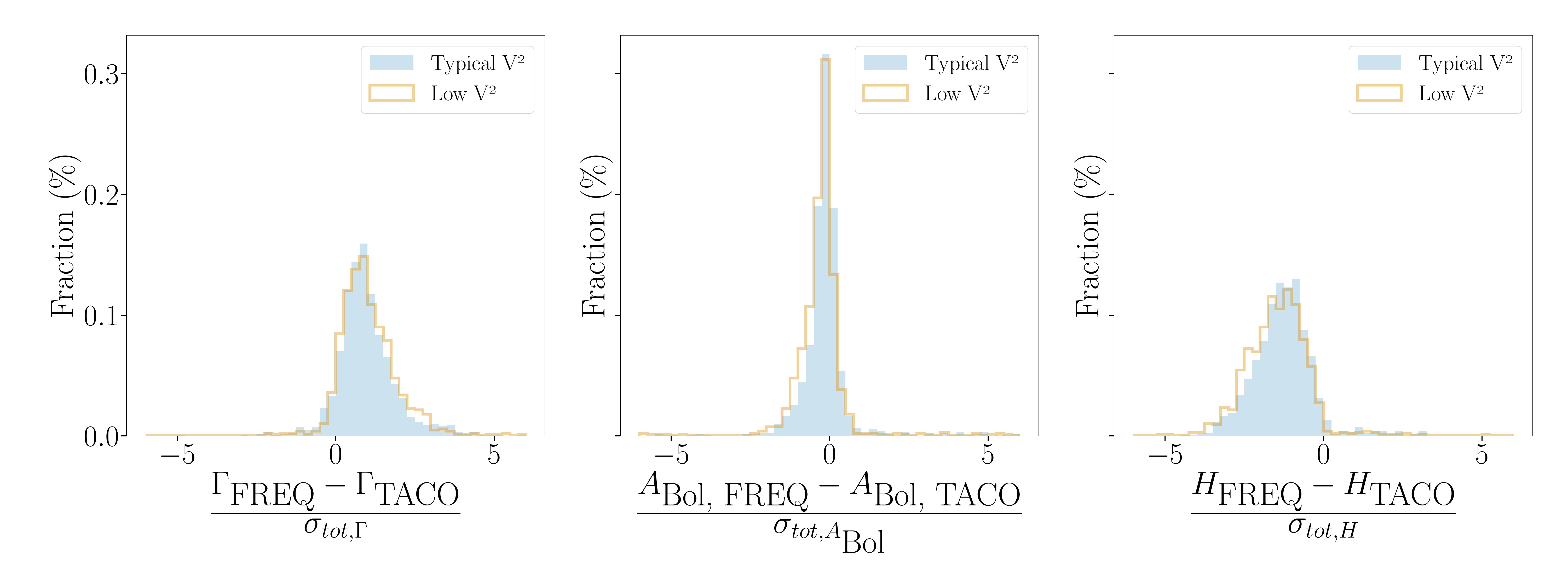}
                \caption{Same as in Fig. \ref{RelDiff_Gamma_ABol_TACO_ABBA_CUM}, but now for the comparison with the FREQ results \citep{2018Vrard}. } \label{RelDiff_Gamma_ABol_TACO_FREQ_CUM}
        \end{figure*}

\begin{figure*}[ht]
                \centering
                \includegraphics[width =0.9\linewidth]{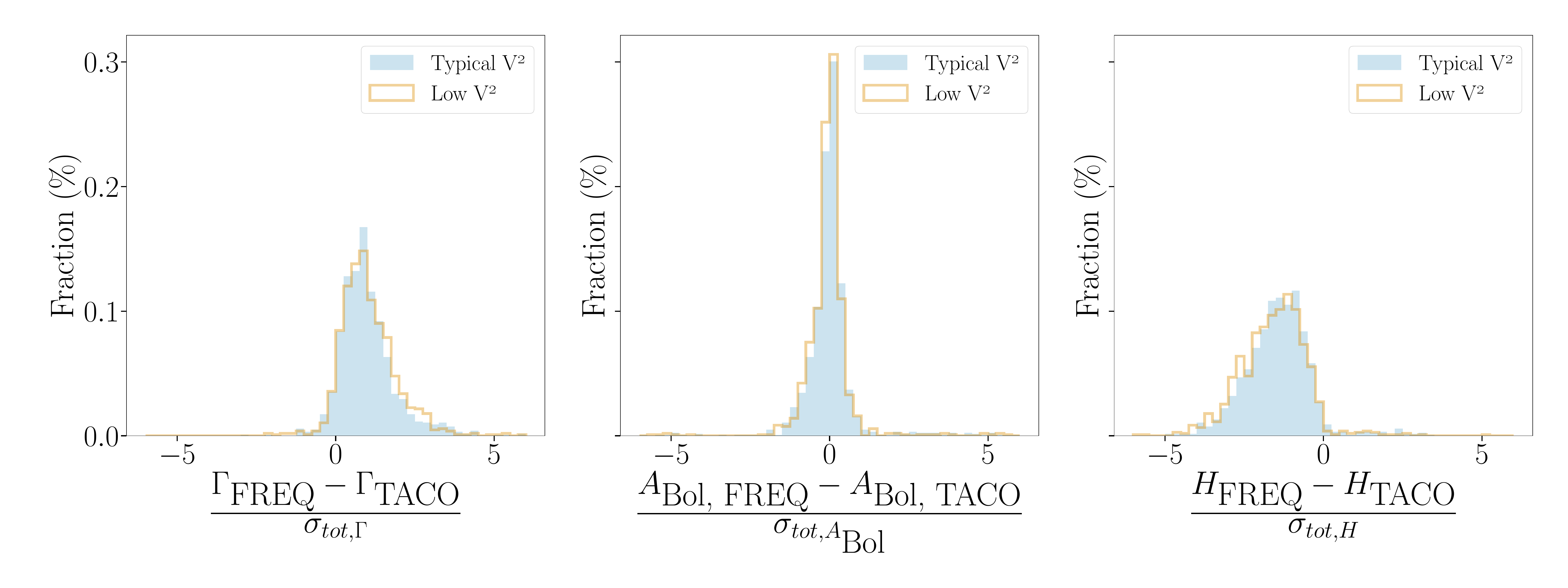}
                \caption{Same as in \ref{RelDiff_Gamma_ABol_TACO_FREQ_CUM}, but now for $S_p$ (blue).}\label{RelDiff_Gamma_ABol_TACO_FREQ_PAIR}
        \end{figure*}

        In this section we compare the values of the linewidths, amplitudes and heights obtained with our code to the values found in the literature. \citet{2019Kallinger} and \citet{2018Vrard} do not report values for the heights of the detected peaks. We therefore computed the heights using Eq.~\ref{ModeAmplitude} with the reported values for the linewidths and amplitudes. We find that our results are in agreement with the values from \citet[see our Figs. \ref{RelDiff_Gamma_ABol_TACO_ABBA_CUM} and \ref{RelDiff_Gamma_ABol_TACO_ABBA_PAIR}]{2019Kallinger}. By comparing our results of our low-visibility sample and our control samples with the values of \citet[see our Figs. \ref{RelDiff_Gamma_ABol_TACO_FREQ_CUM} and \ref{RelDiff_Gamma_ABol_TACO_FREQ_PAIR}]{2018Vrard}, we find that our linewidths are about 40$\%$ narrower than their linewidths (i.e. there is an offset between the two sets of parameters). A similar offset is as expected also observed in the mode heights. These values for the linewidths and heights are, however, still within 3$\sigma$, three times the combined uncertainty in quadrature. By comparing the linewidths obtained by \citet{2018Vrard} and \citet{2019Kallinger}, we also observe that the linewidths in \citet{2018Vrard} are systematically broader. Our results are overall in agreement with the values from \citet{2018Vrard} for the stars present in both samples. Our distributions of our radial-mode parameters are therefore consistent with the distributions obtained with the values of \citet{2018Vrard} and \citet{2019Kallinger} for the stars we have in common.

\newpage

        \end{appendix}
\end{document}